\documentclass[prl,aps,showpacs,amssymb,twocolumn]{revtex4-1}
\usepackage{amsmath}
\usepackage{amssymb}
\usepackage{amsthm}
\usepackage{amsfonts}
\usepackage{listings}
\usepackage{enumerate}
\usepackage{latexsym}
\usepackage{color}
\usepackage[colorlinks=true,urlcolor=blue,citecolor=blue,linkcolor=blue]{hyperref}
\usepackage{psfrag}

\usepackage{bm}
\usepackage{graphicx}
\usepackage{subfigure}

\renewcommand{\vec}[1]{\ensuremath{\mathbf{#1}}}

\newcommand{\beq}{\begin{equation}}
\newcommand{\eneq}{\end{equation}}

\newcommand{\bal}{\begin{align}}
\newcommand{\eal}{\end{align}}

\input{epsf}

\begin{document}

\tolerance 10000

\newcommand{\vk}{{\bf k}}

\title{MoTe$_2$: A Type-II Weyl Topological Metal}

\author{Zhijun Wang$^{1}$}
\author{Dominik Gresch$^{2}$}
\thanks{Zhijun Wang and Dominik Gresch contributed equally to this work}
\author{Alexey A. Soluyanov$^{2}$}
\author{Weiwei Xie$^{3}$}
\author{S. Kushwaha$^{3}$}
\author{Xi Dai$^4$}
\author{Matthias Troyer$^{2}$}
\author{Robert J. Cava$^3$}
\author{B. Andrei Bernevig$^{1}$}
\affiliation{${^1}$Department of Physics, Princeton University, Princeton, NJ 08544, USA}
\affiliation{${^2}$Theoretical Physics and Station Q Zurich, ETH Zurich, 8093 Zurich, Switzerland}
\affiliation{${^3}$Department of Chemistry, Princeton University, Princeton, NJ 08540, USA}
\affiliation{${^4}$Institute of Physics, Chinese Academy of Sciences, Beijing 100190, China}

\date{\today}
\begin{abstract} 
Based on the {\it ab initio} calculations, we show that MoTe$_2$, in its low-temperature orthorhombic structure characterized by an X-ray diffraction study at 100~K, realizes 4 type-II Weyl points between the $N$-th and $N$+1-th bands, where $N$ is the total number of valence electrons per unit cell. Other WPs and nodal lines between different other bands also appear close to the Fermi level due to a complex topological band structure. We predict a series of strain-driven topological phase transitions in this compound, opening a wide range of possible experimental realizations of different topological semimetal phases. Crucially, with no strain, the number of observable surface Fermi arcs in this material is $2$ -- the smallest number of arcs consistent with time-reversal symmetry.
\end{abstract}

\maketitle

The ability of gapless band structures to host topological features was first discussed in the context of liquid He~\cite{Volovik-book, Volovik-Nature97}. It recently became relevant to crystalline materials with the experimental discovery~\cite{Lv-PRX15, Xu-Science15} of the theoretically predicted ~\cite{Huang-NatComm15, Weng-PRX15} Weyl semimetals (WSM) in the TaAs family of compounds. In WSMs a topologically protected band crossing of two bands occurs in the close vicinity of the Fermi level forming a gapless node~\cite{Nielsen-PLB83, Murakami-NJP07, Wan-PRB11}. The low-energy Hamiltonian for such semimetals is that of a Weyl fermion~\cite{Weyl1929}, which exhibits interesting spectroscopic and transport phenomena such as Fermi arcs~\cite{Wan-PRB11, Silaev-PRB12} and the chiral anomaly~\cite{Nielsen-PLB83, Volovik-JETPL86, Xiong-arx15, Zhang-arx15, Huang-PRX15}. 

It was recently shown~\cite{Soluyanov-ARX15} that in materials Weyl fermions come in two flavors: while the type-I Weyl point (WP) (the condensed matter counterpart of the high-energy theory Weyl fermion) is associated with a closed point-like Fermi surface, its newly proposed type-II cousin~\cite{Soluyanov-ARX15} appears at the boundary of electron and hole pockets, and has transport properties that are very different from those of the usual, type-I WSM. Another kind of topological metal -- a nodal line metal~\cite{Burkov-PRB11, Heikkila-JETPL11, Lu-NatPhot13, Weng-PRB15, Kim-PRL15, Xie-APLMat15, Bian-arx15, Fang-PRB15, Yu-PRL15} -- occurs when bands cross along a line in the Brillouin zone (BZ), giving rise to surface states shaped like the surface of a drum~\cite{Yu-PRL15}. The existence of such a nodal line requires the presence of a symmetry, such as mirror symmetry (or the combination of time reversal and inversion in the absence of spin-orbit coupling (SOC)~\cite{generalroute} ), in the material. 

A WP is associated with a topological charge, since it represents  a sink or source of Berry curvature. A nodal line is associated with a Berry phase of $\pi$ along any mirror-symmetric closed trajectory linking with the line. According to the fermion doubling theorem~\cite{Nielsen-NPhB81, Nielsen-PLB81} the number of sinks in a crystal has to be equal to the number of sources, meaning that WPs can only appear and annihilate in pairs of opposite topological charge. In non-magnetic materials the presence of time-reversal symmetry dictates the minimal number of WPs to be four, giving rise to two Fermi arcs on the surface of the material. The WSMs experimentally discovered~\cite{Lv-PRX15, Xu-Science15, Shekhar-arx15} and theoretically predicted~\cite{Huang-NatComm15, Weng-PRX15, Liu-PRB14, Hirayama-PRL15, Bzdusek-PRB15} to date all have more than the minimal number of WPs, as well as a multitude of Fermi arcs, which prevent clean spectroscopy. Of type-I WSMs the TaAs family~\cite{Huang-NatComm15, Weng-PRX15}, hosts 24 WPs and the recently predicted type-II WSM WTe$_2$ hosts 8 of them between the $N$-th and $N+1$-th bands, where $N$ is the total number of valence electrons per unit cell. Below we refer to the bands below band $N$ inclusive as valence, and the ones above as conduction.

In search for other type-II WSMs, it is natural to look at compounds chemically similar to WTe$_2$. One such compound, MoTe$_2$ in a previously unreported orthorhombic phase was argued to be a strong candidate for another realization of type-II WSM~\cite{Sun-PRB15, Soluyanov-ARX15}.
A very recent interesting work~\cite{Sun-PRB15} reported that orthorhombic MoTe$_2$ also hosts $8$ type-II WPs in the $k_z=0$ plane between bands $N$ and $N+1$ (as  WTe$_2$), for the MoTe$_2$ crystal structure measured at 120~K. By further analyzing the MoTe$_2$ structure reported in Ref.~\cite{Sun-PRB15}, we found that in addition there are 16 WPs out of the $k_z= 0$ plane also formed by bands $N$ and $N+1$, located in the immediate vicinity of the Fermi level~\footnote{Two of these points are at ${\bf k}=(0.12661133,  0.0996582 ,  0.21142578)$ ($20$~meV above $E_{\mathrm{F}}$) and ${\bf k}=(0.11034922,  0.07301836,  0.14727305)$  ($36$~meV above $E_{\mathrm{F}}$). Another 6 points are related to these ones by the mirror reflections $M_x$ and $M_y$, and 8 more points are located symmetrically to these ones about the $k_z=0$ plane}. 


In this paper we present the experimental structure of MoTe$_2$ at 100~K and use it to perform first principles and tight-binding calculations of the band structure topology around the Fermi level. Our calculation suggests a different topological physics around the Fermi level than that reported in Ref.~\cite{Sun-PRB15}: we find only $4$ type-II WPs (which we call $W$) between bands $N$ and $N+1$ $55$~meV above the Fermi level. These WPs give rise to only 2 clean visible Fermi arcs on the surface of this material. In this sense MoTe$_2$ represents a "hydrogen atom" of time-reversal invariant WSMs, having the minimal possible number of WPs consistent with time-reversal symmetry. We  provide arguments that the difference with the $8$ WPs in the $k_z=0$ plane reported in Ref.~\cite{Sun-PRB15} comes from the high sensitivity of the band structure of MoTe$_2$ to even small changes in lattice parameters -- MoTe$_2$ lies on a cusp between a transition from $4$ to $8$ nodes in between valence and conduction bands in the $k_z=0$ plane. Indeed, there are small differences in the crystallographic data of Ref.~\cite{Sun-PRB15} at 120~K and the one reported here at 100~K resulting in the unit cell volume decrease of about 1\%, suggesting the possibility of a temperature-driven topological phase transition. 

Another recent work~\cite{Chang-arx15} predicts the existence of WPs in Mo doped WTe$_2$. That prediction is obtained by interpolating between the tight-binding models of WTe$_2$ and a theoretically relaxed orthorhombic MoTe$_2$. Such interpolation represents a very strong approximation for the band structure of a doped compound, which, together with the above discussed sensitivity of the WPs to even small differences in the experimental crystal structure, makes the predictions of the work~\cite{Chang-arx15} unreliable. More valuable discussions about strain-driven topological phase transition are presented in the Supplementary Information (S.I.). 


Moreover, in metals it is important to look at other topological features near the Fermi level, not only those formed between valence and conduction bands, since the occupation becomes a function of crystal momentum $k$ in this case~\cite{Gosalbez-PRB15}, and we find many such additional features in MoTe$_2$. Inspection of crossings between bands other than $N$ and $N+1$, occurring close to the Fermi level, reveals many additional topologically protected crossings formed by the conduction bands $N+1$ and $N+2$, including line nodes and WPs, some of which (20 in total) are close in energy to the $W$ points (see S.I.).  We find also two nodal lines close to the Fermi level, formed by the valence bands $N-1$ and $N$, protected by mirror symmetry. Unlike $W$ points that are formed at the boundary of electron and hole pockets, the additional topological features arise at the touching points of two pockets of the same carriers. Thus, despite a complex topological band structure, the surface Fermi arcs arising due to the four type-II $W$ points are rather clean and should be easy to see in spectroscopic experiments, while the surface states associated with the additional topological crossings overlap with surface projections of the bulk states. 
\begin{figure}[tb]
\centering
\includegraphics[width=8 cm]{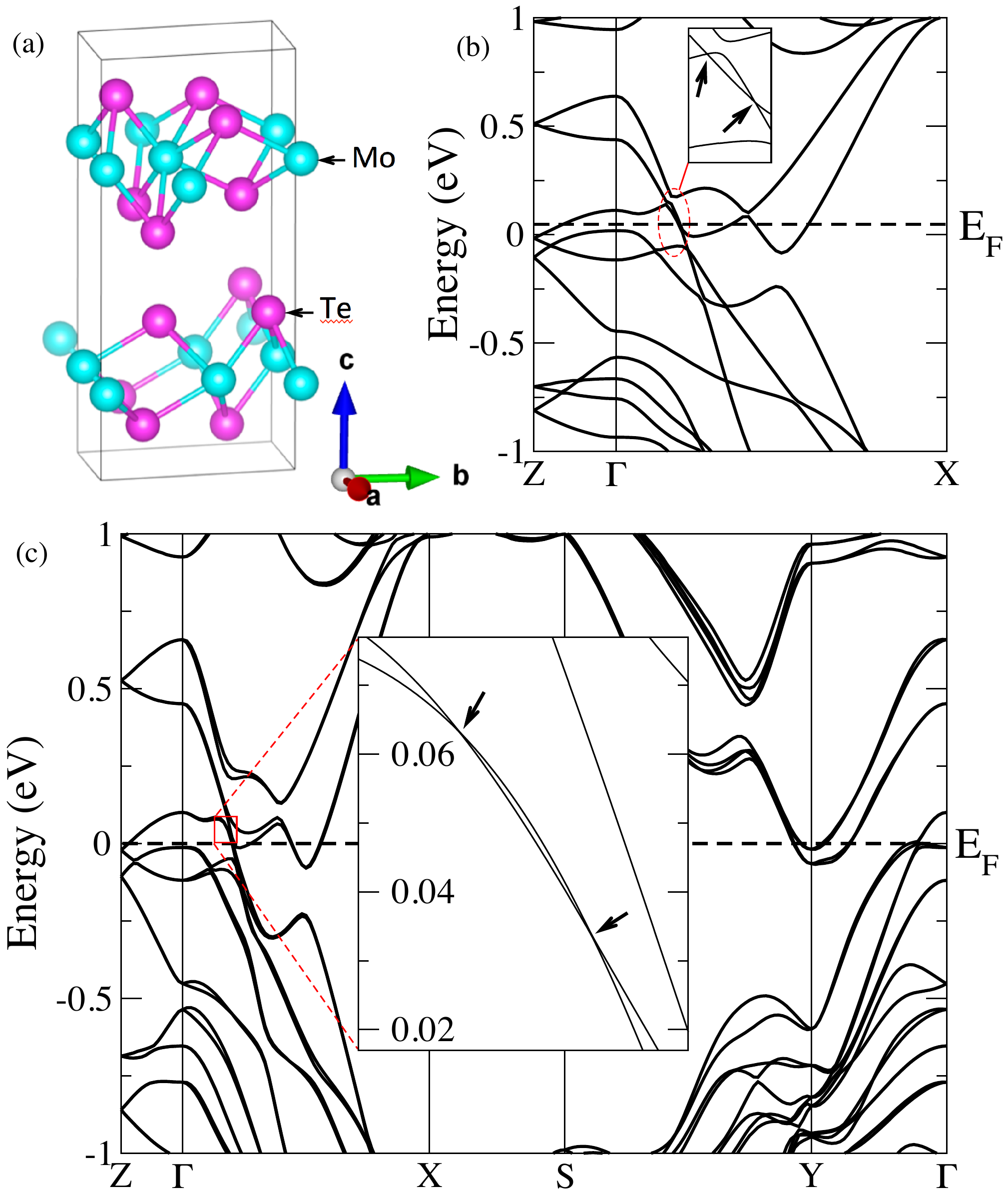}
\caption{Atomic and electronic structure of MoTe$_2$. (a) The orthorhombic crystal structure of Td-MoTe$_2$ with Te atoms forming a distorted octahedron around Mo atoms. (b) The band structure of MoTe$_2$ in the absence of spin-orbit coupling. The arrows indicate two crossing points as part of the nodal line. (c) The band structure of MoTe$_2$ with spin-orbit coupling. The inset shows crossings between the $N-1$-th and $N$-th bands which are part of a nodal line.}
\label{str}
\end{figure}

We grew samples by slow cooling and performed diffraction measurements at 100~K. Our results establish that MoTe$_2$ has a new low temperature orthorhombic 1T'-phase (as was previously reported in \cite{Sun-PRB15}), which we designate as the $\gamma$-phase (see S.I. for the full structure characterization). Using this crystallographic data, we perform {\it ab initio} calculations based on the density functional theory (DFT)~\cite{Hohenberg-PR64,Kohn-PR65} and the generalized gradient approximation (GGA) for the exchange-correlation potential~\cite{PBE}. We first compute the band structure of MoTe$_2$ without spin-orbit coupling (SOC), as illustrated in Fig.~\ref{str}(b). We find 2 mirror-protected nodal lines in the k$_y$=0 plane (more details in S.I.) and 12 WPs formed by valence and conduction bands, 4 of which are located in the $k_z=0$ plane (W1 points) and 8 are out of that plane (W2 points), as shown in Tab.~\ref{posi}.

\begin{figure}[htb]
\centering
\includegraphics[width=8 cm]{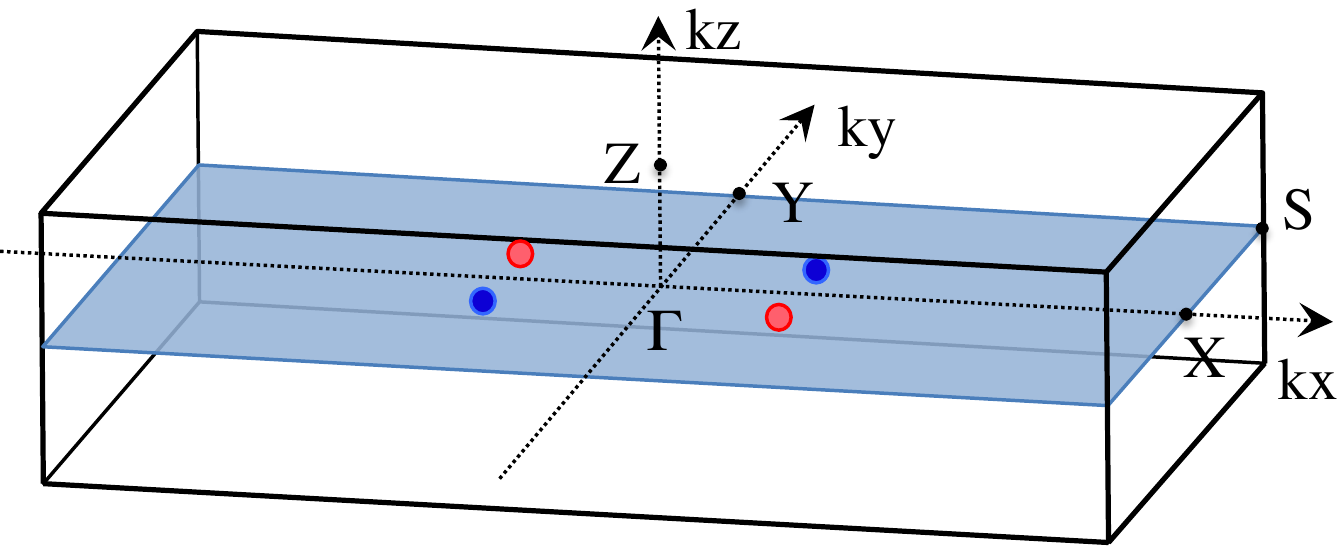}
\caption{Weyl points in the Brillouin zone of MoTe$_2$. Four Weyl points formed by bands $N$ and $N+1$ in the Brillouin zone are shown. The blue and red colors indicate Chern numbers +1 and -1 respectively.}
\label{BZ}
\end{figure}

The strong SOC of Mo 4$d$-  and Te 5$p$-states which dominate the physics around $E_{\mathrm{F}}$ significantly changes the band structure as shown in Fig.~\ref{str}(c). We first elucidate the topological crossings between valence and conduction bands. The two nodal loops present without SOC become fully gapped. The structure of WPs also changes significantly: the WPs at $k_z\neq 0$ disappear, while only four WPs are found in the $k_z=0$ plane, still allowed by the $C_{2T}$ symmetry~\cite{Soluyanov-ARX15, generalroute} (see symmetries in S.I.). The coordinates of these points ($W$) are given in Tab.~\ref{posi}. Their location and Chern numbers (see S.I. for the details of Chern number calculation) are illustrated in Fig.~\ref{BZ}. The separation between the nearest points with opposite Chern numbers in the unstrained MoTe$_2$ is  $\approx$10\% of the reciprocal lattice constants $|G_y|$ meaning that the topological Fermi arcs should be easily observable in this material. While, as shown below, other topological gapless features show up very close to the $E_{\mathrm F}$ in between bands other than $N$ and $N+1$, it is the $W$ WPs that are most important as they give rise to the only Fermi arcs not superimposed on bulk states upon surface projection. 
\begin{table}[tb]
\begin{center}
\caption{WPs of MoTe$_2$. The positions (in reduced coordinates $k_x$, $k_y$, $k_z$), Chern numbers, and the energy relative to the $E_{\mathrm{F}}$ are given. W1 and W2 are the WPs formed by bands $N/2$ and $N/2+1$ in the absence of SOC, while $W$ are the WPs formed by bands $N$ and $N+1$ with full SOC . The coordinates of the other points are related to the ones listed by the reflections $M_{x,y}$.}
\label{posi}
\begin{tabular}{cccc}
\hline
\hline
Weyl points & Coordinates & Chern number & $E-E_{\mathrm{F}}$ \\
& ($k_x\frac{2\pi}{a},k_y\frac{2\pi}{b},k_z\frac{2\pi}{c}$) &  & (meV) \\
\hline
W1 & $(0.1819,0.1721,0)$ &$+1$ &$-38$ \\
W2 & $(0.1300,0.0793,\pm0.298)$ &$-1$ &$-18$ \\
\hline
$W$ & $(0.1011,0.0503,0)$ &$+1$ &$+55$\\
\hline
\hline
\end{tabular}
\end{center}
\end{table}

We computed topological invariants to establish the existence of the WPs, and to prove that no additional WPs are present in between the valence and conduction bands. The first  invariant is the Chern number, associated with each of the WPs, which was computed  both from the Wannier-based tight-binding model~\cite{Souza-PRB01, wannier90}, and directly from first-principles calculations~\cite{Z2Pack} (see S.I.). The result of this calculation is illustrated in Tab.~\ref{posi} and Fig.~\ref{BZ}, where the WPs and their Chern numbers are shown in the BZ. 

\begin{figure}[tbh]
\centering
\includegraphics[width=8 cm]{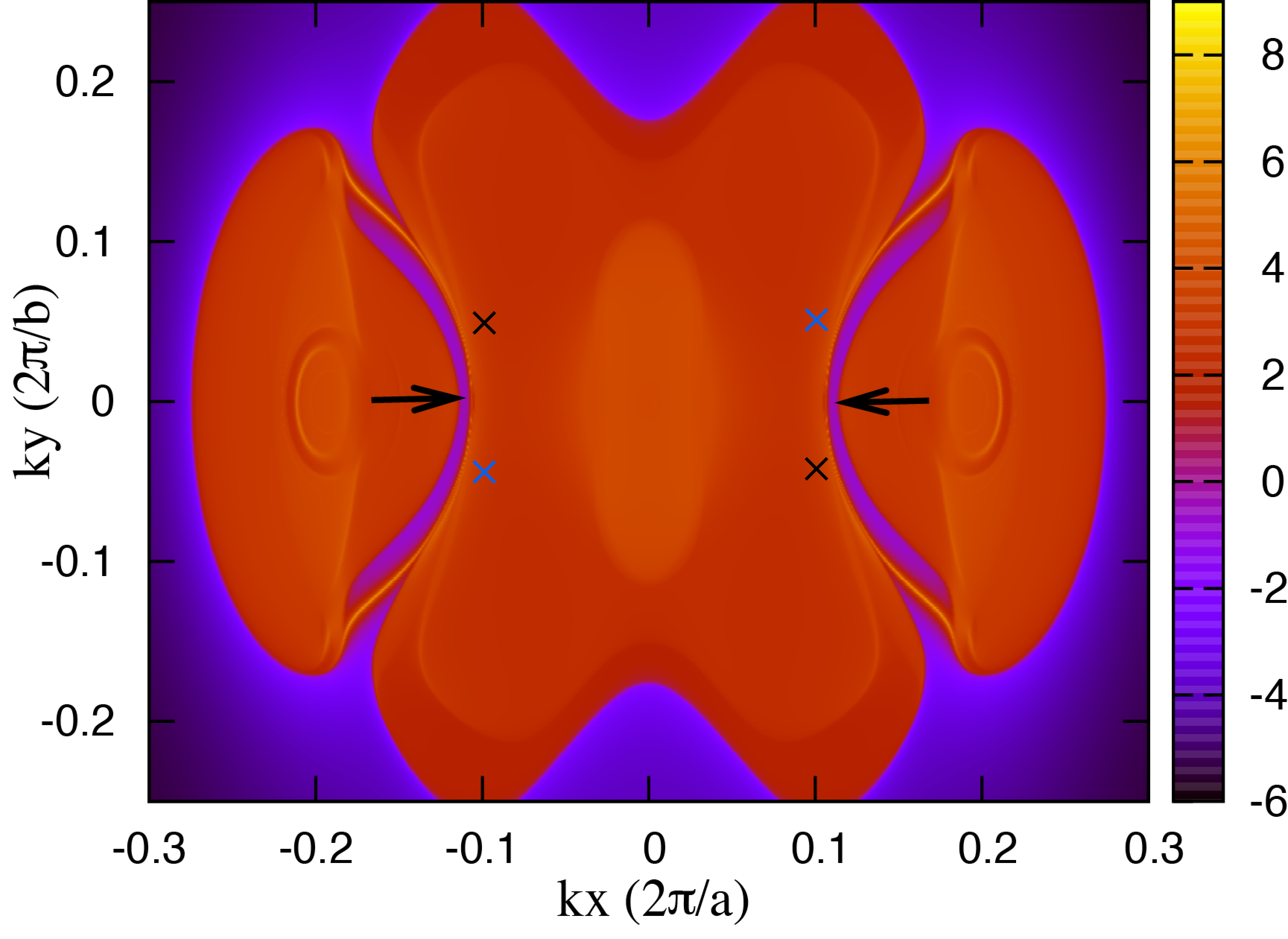}
\caption{Spectral function of the $(001)$-surface of the orthorhombic MoTe$_2$ at $20$~meV below the Fermi level. A projection of a big hole-like pocke is in the center, with the projections of the electron-pockets to the left and to the right from it. In the gap of these two kinds of pockets, topological Fermi arcs are marked with arrows. The WPs' projections are denoted by x, with blue and black colors indicating Chern numbers +1 and -1 respectively.}
\label{wps}
\end{figure}

Further insight into the topology of the Bloch bands of MoTe$_2$ is obtained by computing the $\mathbb{Z}_2$ invariant of the lowest $N$ bands on the time-reversal symmetric planes $k_i=0,\pi$. The $k_z=0$ plane contains the WPs and is gapless; hence no such invariant can be defined on this plane. The other five planes, however, are gapped between bands $N$ and $N+1$. Of these planes only the $k_y=0$ one has a non-trivial $\mathbb{Z}_2$ invariant. This means that the $k_y=0$ cut of the BZ is analogous to the 2D BZ of a quantum spin Hall insulator that carries an odd number of Kramers pairs of edge states. The lack of a nontrivial $\mathbb{Z}_2$ invariant on all the other planes implies the existence of disconnected Fermi surfaces. Notice that a connected Fermi sea of the surface states, be it strong or weak topological or trivial insulator, does not lead to only one nontrivial $\mathbb{Z}_2$ index on a high symmetry plane. Fig.~\ref{wps} shows the surface spectral function for the $(001)$-surface of MoTe$_2$, and topological Fermi arcs crossing the $k_y=0$ plane are clearly visible. 

In type-I WSMs tuning the Fermi level to the energy of the WP results in the surface Fermi arcs connecting projections of the WPs on the particular surface. Type-II WPs appear at the boundary between the pockets, when  $E_{\mathrm{F}}$ is tuned to the WP energy; hence  projections of bulk carrier pockets necessarily appear in the surface electronic density of states irrespective of the Fermi energy in relation to the WP. The Fermi arcs can in this case be hidden within the projection of the bulk pockets on the surface, but they can still be revealed by tuning the chemical potential (see S.I.). This is illustrated for MoTe$_2$ in Fig.~\ref{wps}, where we chose a spectroscopically reachable value for the chemical potential of $-20$~meV below $E_\mathrm{F}$. The clean Fermi arcs have been revealed in the angle-resolved photoemission spectroscopy measurement~\cite{nan_mote2}. A projection of a big hole-like bulk pocket is seen in the center of the surface BZ, with the projections of the electron-pockets to the left and to the right from it. Unlike the case of type-I WSMs, where a Fermi arc connects surface projections of WPs of opposite chirality, the Fermi arcs illustrated in Fig.~\ref{wps} are of different nature. 

The projections of the $W$ WPs are within the projected hole pocket, and all the Fermi pockets have zero Chern numbers, so that in general no Fermi arcs connecting different pockets should appear. However, any $(k_x,k_z)$ cut of the BZ in between two adjacent $W$ points ($|k_y|<0.0503$) has to exhibit the quantum spin Hall effect, meaning that a Kramers pair of surface states connecting valence and conduction states has to appear in the gap between them, resulting in topological Fermi arcs. Since the arc cannot appear without being connected to projections of WPs or carrier pockets, for $|k_y|>0.0503$ it is the topologically trivial state that continues the arc to merge it into the electron pocket (see S.I. for illustrations). The resultant two Fermi arcs represents the cleanest observable consequence of the type-II WPs. 

\begin{figure}[tb]
\centering
\includegraphics[width=8 cm]{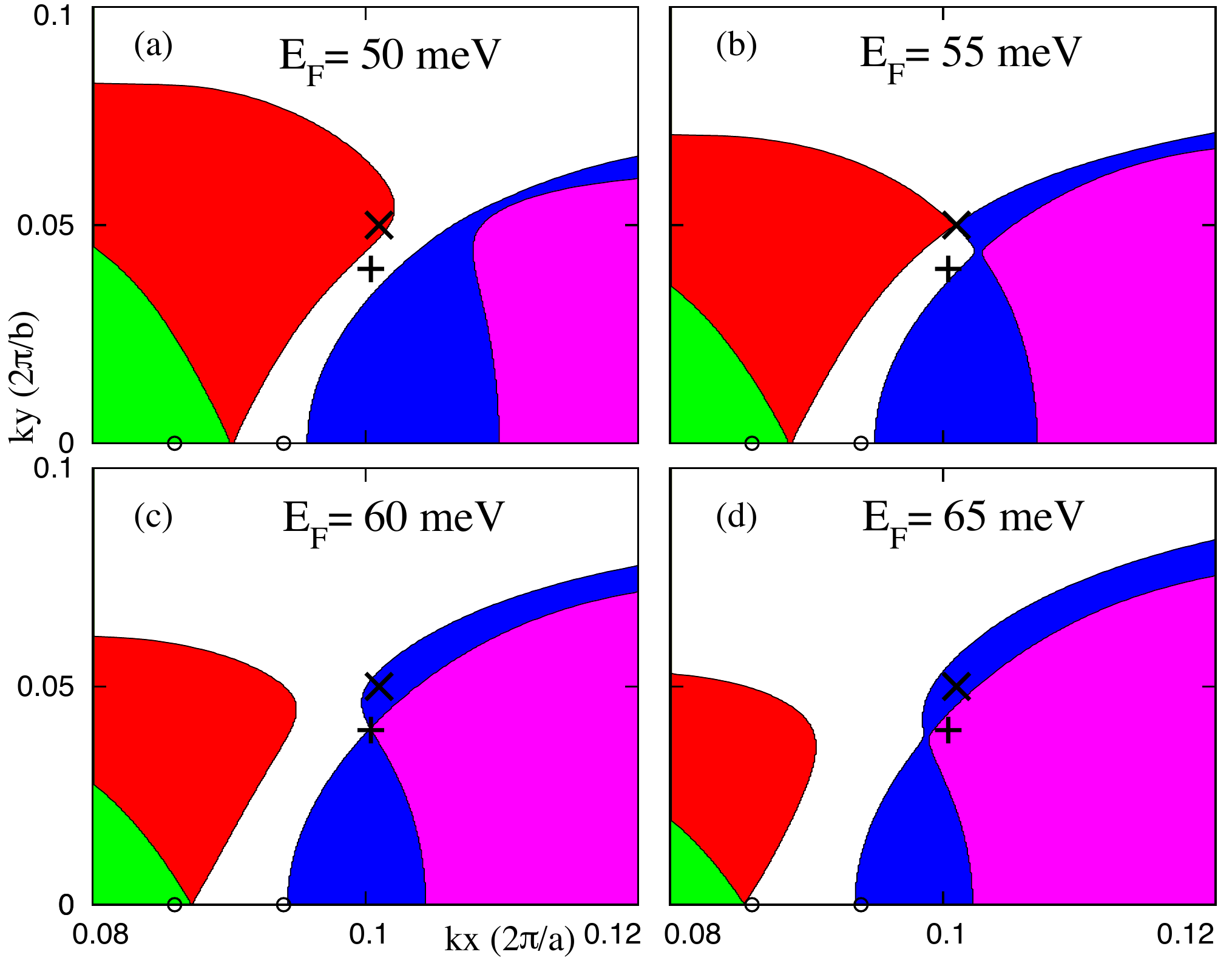}
\caption{Contour plot of the bulk Fermi surface at $k_z=0$ for different values of $E_{\mathrm{F}}$. The Weyl point $W$ between bands $N$ and $N+1$ is designated with a $\times$ sign, while the Weyl point between bands $N+1$ and $N+2$ is shown as a $+$ sign. The circles mark the crossings of the plane with a nodal line formed by bands $N-1$ and $N$. Pockets formed by bands $N-1$, $N$, $N+1$ and $N+2$ are shown in green, red, blue and cyan correspondingly.}
\label{bulkFS}
\end{figure}

A cross section of the bulk Fermi surface in the $k_z=0$ plane around the $W$ points is illustrated in Fig.~\ref{bulkFS}. The Fermi surface consists of two hole (p) and four electron (n)  pockets. The latter form two pairs located to the left (not shown) and to the right (shown) of the p-pockets. When $E_\mathrm{F}$ is below the position of the $W$ (panel (a)) p- and  n-pockets come in pairs of interpenetrating sheets. All the $W$ WPs are inside the p-pocket.  Upon increasing $E_\mathrm{F}$ the p- and n-pockets approach each other and eventually touch at $W$ (panel b). Further increase of $E_\mathrm{F}$ splits the pockets again, but now $W$ is inside the n-pockets (panels c and d). At all times all the pockets have zero Chern number, enclosing an equal number of WPs with opposite chiralities.

While this concludes the analysis of the WPs between the bands $N$ and $N+1$, MoTe$_2$ exhibits several other topologically protected crossings close to the Fermi level. Even though none of these topological features impact the surface states spectroscopy shown in Fig.~\ref{wps}, as they project inside bulk bands on the surface, we analyze them here for completeness. At only $2$ and $5$~ meV above $W$, two new quartets of WPs exists between the bands $N+1$ and $N+2$ in the $k_z=0$ plane. One of these points is clearly seen in Fig.~\ref{bulkFS}(c) occurring at the touching points of the two n-pockets. Calculation of additional topological invariants further confirmed the existence of these WPs.

Furthermore, the inset of Fig.~\ref{str}(c) shows crossings between bands $N$ and $N-1$ occurring on the high symmetry line $\Gamma X$. Symmetry considerations~\cite{generalroute} indicate that such crossings on a high-symmetry plane can only occur if a line node is present in the mirror plane $k_y=0$ of the BZ. Indeed, we find that these bands have different glide plane $M_y$ eigenvalues and two line nodes exist in the $k_y=0$ plane related by mirror $M_x$. A further check of the topological nature of this ring is obtained by computing the Berry phase of a loop trajectory that links with the nodal line. We find this to be equal to $\pi$, as expected for a nodal line.

In conclusion, based on the DFT calculations, we have studied in detail the topological properties of a new orthorhombic $\gamma$-phase of MoTe$_2$, which has been experimentally characterized. The {\it ab initio} calculations suggest that unlike the $\alpha$ and $\beta$ phases, the $\gamma$-phase hosts a multitude of topological features around the Fermi level including type-II WPs and nodal lines. We found that the WPs between the $N$-th and $N+1$-th come in a single quadruplet, the smallest number allowed by time-reversal symmetry. This allows for a particularly clean Fermi arc structure on the surface of MoTe$_2$, which should be readily observable in spectroscopic measurements~\cite{nan_mote2}. Other WPs and nodal lines between different other bands also appear in MoTe$_2$, but their spectroscopic signatures on the surface overlap with those of the projected bulk Fermi surfaces. 

\noindent \textbf{Acknowledgments} We thank Binghai Yan  for helpful discussions.  This work was supported by NSF CAREER DMR-095242, ONR - N00014\text{-}14\text{-}1-0330, ARO MURI W911NF-12-1-0461,  NSF-MRSEC DMR-0819860, Packard Foundation and Keck grant. D.G., A.S. and M.T. were supported by Microsoft Research, the European Research Council through ERC Advanced Grant SIMCOFE, the Swiss National Science Foundation through the National Competence Centers in Research MARVEL and QSIT. Z.W. and X.D. were supported by the National Natural Science Foundation of China (No. 11504117), the 973 program of China (No. 2013CB921700), and the ``trategic Priority Research Program (B)'' of the Chinese Academy of Sciences (No. XDB07020100).

\bibliography{paper}

\section*{SUPPLEMENTARY INFORMATION}

\appendix

\section{Crystal structure of MoTe$_2$}
To obtain the low temperature crystal structure of MoTe$_2$ we grew samples by slow cooling and performed diffraction measurements at 100~K to obtain the lattice constants, lattice space group, and atomic coordinates. 
%
\subsection{The crystal structure of orthorhombic MoTe$_2$}
The crystal structure of orthorhombic MoTe$_2$, the 1T' form, was determined experimentally at 100 K by single crystal X-ray diffraction. This is a so far (with the exception of the very recent paper of Ref\cite{Sun-PRB15}) uncharacterized structure for MoTe$_2$, whose crystal structure has been reported previously for a hexagonal symmetry phase (the $\alpha$ form), 2H MoTe$_2$, and a monoclinic symmetry phase (the $\beta$ form), 1T'' MoTe$_2$~\cite{brownbeta, Vellinga-1970, Ikeura-2015}. The crystal structure of the previously reported monoclinic form is related to the structure of orthorhombic WTe$_2$~\cite{brownbeta}, but is distorted and therefore is not isostructural with it. Thus, the orthorhombic 1T' form of MoTe$_2$ characterized here, isostructural with orthorhombic WTe$_2$, is the third characterized structural variant of MoTe$_2$ known; it therefore can be designated alternatively as the $\gamma$ phase.

\subsection{Experimental details}
The crystals of orthorhombic MoTe$_2$ were made by slow cooling (at 1.5~K/hr) a Te-rich flux ($\sim$95~\% Te) from 1000~C to 820~C and then centrifuging off the flux. They were then annealed in a sealed evacuated quartz tube for $\sim$12 hours in a thermal gradient, with the crystals at 400~C and the cold end of the tube at about 60~C.

Single-crystal data were collected at 100~K on a Bruker Apex II diffractometer with Mo K$\alpha_1$ (=0.71073~\AA) radiation. Data were collected over a full sphere of reciprocal space with 0.5${}^\circ$ scans in $\omega$ with an exposure time of 30~s per frame. The SMART software was used for data acquisition. Intensities were extracted and corrected for Lorentz and polarization effects with the SAINT program. Numerical absorption corrections were accomplished with XPREP which is based on face-indexed absorption~\cite{SHELXTL}. With the SHELXTL package, the crystal structure was solved using direct methods and refined by full-matrix least-squares on F${}^2$~\cite{Sheldrick}. The largest peak in the final $\Delta$F map was 4.91~e\AA${}^{-3}$, 1.03~\AA from Te1, and the largest hole was -3.94~e\AA${}^{-3}$, located 0.98~\AA from Mo1. The crystal refinement and atomic parameters are given in Tabs.~\ref{tab1}-\ref{tab2}.
\begin{table}[htb]
\begin{center}
\caption{Single crystal crystallographic data for orthorhombic MoTe$_2$ at 100~K. This is the 1T' or $\gamma$ form.}
\label{tab1}
\begin{tabular}{cc}
\hline
Refined Formula & MoTe$_2$\\
F.W. (g/mol)&  351.14 \\
Space group; $Z$&  $P_{mn2_1}$(No.31); 4\\
a(\AA)& 3.4582(10)\\
b(\AA)& 6.3043(18)\\
c(\AA)& 13.859(4)\\
V(\AA${}^3$)& 302.1(2)\\
Absorption Correction & Numerical\\
Extinction Coefficient & None\\
$\theta$ range (deg)& 3.55-29.566\\
No. reflections & 2765\\
No. independent reflections & 899\\
No. parameters & 38\\
$R_1$; $wR_2$ (all $I$) & 0.0579; 0.1223\\
Goodness of fit & 1.004\\
Diffraction peak and hole (e-/\AA${}^3$)& 4.913; –3.941\\
\hline
\end{tabular}
\end{center}
\end{table}
\begin{table}[htb]
\begin{center}
\caption{Atomic coordinates and equivalent isotropic displacement parameters for orthorhombic MoTe$_2$ at 100~K. Noncentrosymmetric space group $P_{mn2_1}$. This is the 1T' or $\gamma$ form.  $U_{eq}$ is defined as one-third of the trace of the orthogonalized $U_{ij}$ tensor (\AA${}^2$).}
\label{tab2}
\begin{tabular}{ccccccc}
\hline
Atom & Wyckoff. & Occupancy & $x$ & $y$& $z$& $U_{eq}$\\
\hline
Mo(1)& 2$a$&  1& 0& 0.0297(6)& 0.7384(6)& 0.012(2)\\
Mo(2)& 2$a$&  1& 0& 0.6062(7)& 0.2240(7)& 0.014(2)\\
Te(1)&  2$a$&  1& 0& 0.2163(8)& 0.1269(5)& 0.015(2)\\
Te(2)&  2$a$&  1& 0& 0.6401(8)& 0.8363(5)& 0.012(2)\\
Te(3)&  2$a$&  1& 0.5& 0.1374(8)& 0.8783(5)& 0.014(2)\\
Te(4)&  2$a$&  1& 0.5& 0.7090(9)& 0.0827(5)& 0.010(2)\\
\hline
\end{tabular}
\end{center}
\end{table}
\subsection{ Symmetries of the $\gamma$-phase }
The crystal structure of $\gamma$-MoTe$_2$ with 4 formula units in the unit cell is shown in Fig.~(1) in the main text.  The corresponding point group is $C_{2v}$ and there are three symmetry operations: the symmorphic reflection $M_x$, the non-symmorphic reflection $M_y$, and the non-symmorphic $C_{2z} = M_x M_y$ rotation. The translation accompanying the two non-symmorphic operations is $(0.5,0,0.5)$ in units of the lattice constants. Crucially, in the $k_z=0$ plane a little group exists at each point $(k_x, k_y, 0)$ formed by the product of time-reversal and $C_{2z}$, $C_{2T}=T*C_{2z}$. It is shown in Ref.~\cite{Soluyanov-ARX15, generalroute} that this symmetry allows (just like in WTe$_2$) for the presence of WPs in the $k_z=0$ plane.  

\section{Details of numerical calculations}

The electronic structure calculations have been carried out using the all-electron WIEN2K package~\cite{w2kcode, wien2k2}. A $15\times8\times3$ mesh and the exchange-correlation functional with a generalized gradient approximation (GGA) parametrized by Perdew, Burke, and Ernzerhof (PBE) have been used~\cite{ggapbe}. 

The results for the electronic band structure and topological invariants were verified versus the pseudopotential calculations done in VASP~\cite{VASP}, using PAW~\cite{PAW1, PAW2} pseudopotentials with $4s^24p^65s^14d^5$ and $5s^25p^4$ valence electron configurations for Mo and Te, respectively. 
Spin-orbit coupling was included in the pseudopotentials, and the PBE approximation~\cite{ggapbe} was used. Self-consistent field calculations were performed on a $16 \times 10 \times 4$ $\Gamma$-centered grid, with a Gaussian smearing of width $0.05$~eV. The energy cut-off was chosen at $450$~eV. Additional calculations were done using a $18 \times 15 \times 9$ $\Gamma$-centered grid, with an energy cut-off of either $260$ or $300$~eV, and also using a Gaussian smearing width of $0.05$~eV. The experimental lattice parameters listed in Tabs.~\ref{tab1}-\ref{tab2} are used in the calculations. To calculate the surface states, Wannier functions based tight-binding models for Mo $4d$ and Te $5p$ orbitals have been constructed~\cite{Souza-PRB01, wannier90}. The topological invariants were verified with both tight-binding and \textit{ab initio} calculations, where the Z2Pack~\cite{Z2Pack} 
(\url{http://z2pack.ethz.ch/doc}) package was used for the latter.

\section{Band structure in the absence of spin-orbit coupling}

The band structure of MoTe$_2$ exhibits several topological features in the absence of spin-orbit coupling (SOC). A clear band inversion and multiple band crossings are found in the band structure around $E_{\mathrm F}$ along the $\Gamma X$ line, which is part of the $k_y=0$ mirror plane. We find that the two bands $N/2$, $N/2+1$ (spin is not taken into account) cross along the $\Gamma X$ line, having opposite glide-plane eigenvalues $M_{y}= \pm e^{-i(k_x+k_z)/2}$. Theoretical symmetry analysis~\cite{generalroute} dictates the appearance of a line node in the $k_y=0$ plane in case of a degeneracy on the $\Gamma X$ line. Indeed, in this plane we find two nodal lines, related by $M_x$. The degeneracy point found on the $\Gamma X$ line belongs to one of these nodal lines. In addition, 12 WPs between the $N/2$ and $N/2+1$ bands are found in MoTe$_2$ in the absence of SOC. Four of these points are located in the $k_z=0$ plane, while the other eight appear off-plane as two quartets, symmetrically located about $k_z=0$. One in-plane point W1 and two out of plane points W2, are listed in Tab.~I of the main text, with the other 9 points related to these by $M_x$ and $M_y$. 

\section{Weyl points between bands $N$ and $N+1$}

\begin{figure}
\includegraphics[width=0.5\columnwidth]{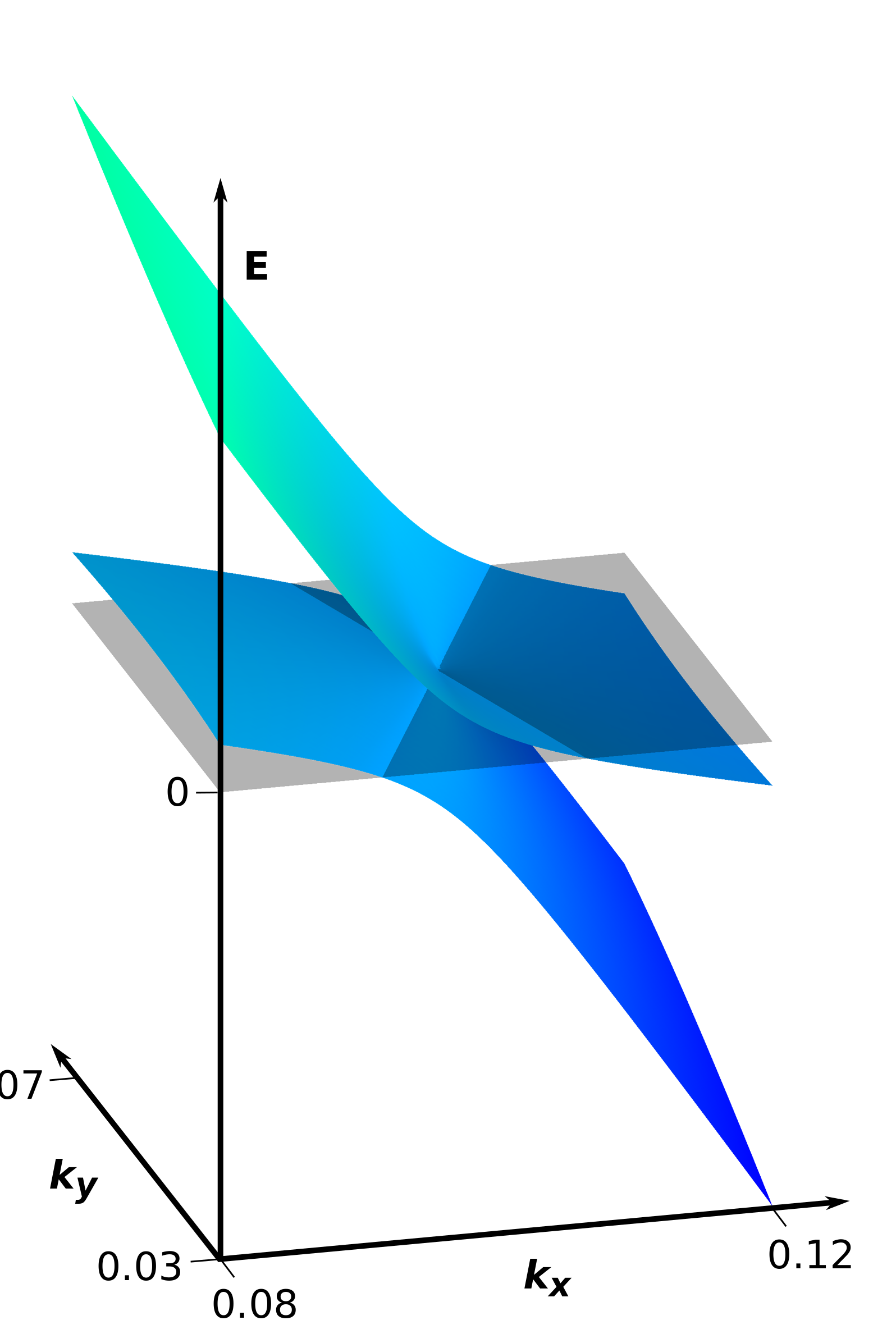}
\caption{Linear fit of the Weyl node $W$ in the $k_z=0$ plane. The plane at zero energy corresponds to the location of $E_{\mathrm{F}}$.}
\label{WP_W3}
\end{figure}

As described in the main text, there are 4 Weyl points (called $W$ in the main text) formed by bands $N$ and $N+1$ in MoTe$_2$ for the structure reported in this paper. The number $N$ corresponds to the number of valence electrons per unit cell. These points are of type-II, as can be seen from the band dispersion obtained from first-principles calculations. The dispersion for the linearized Hamiltonian is illustrated in Fig.~\ref{WP_W3} clearly showing the Weyl point at the boundary between electron and hole pockets.
A general type-II WP~\cite{Soluyanov-ARX15} Hamiltonian is written in terms of the Pauli matrices $\sigma_{x,y,z}$ and a kinetic term described by a unit matrix $I$
\begin{equation}
\mathcal{H}(\vec{k}) =\sum_{i=1}^3 v_i k_i~I + \sum_{i, j=1}^3 k_{i} A_{i, j} \sigma_{j} 
\end{equation}
where there exists a cone of directions in $k$-space, in which the first (kinetic) term of the Hamiltonian dominates over the second (potential) one, that is $(\sum_i v_i k_i)^2>\sum_j(\sum_i k_{i} A_{i, j})^2$. Fitting the theoretical model derived from the symmetry analysis to the band structure around the WPs obtained from {\it ab initio} calculations results in the following effective Hamiltonian for the $W$ WP
\begin{equation}
H({\bf k})=v_1 k_x+v_2 k_y+(ak_x+bk_y)\sigma_y+(ck_x+dk_y)\sigma_z+ek_z\sigma_x
\label{ham}
\end{equation}
with parameters (in eV\AA) $v_1=-3.39$, $v_2=0.58$, $a=0$, $b=0.78$, $c=2.6$, $d=-0.58$ and $e=-0.0045$. Since there is a direction around which the kinetic energy dominates ($\hat{k}\parallel \hat{x}$),  $W$ is a type-II WP. If we limit our analysis to the bands $N$ and $N+1$ forming only $4$ WPs, then MoTe$_2$ would be the simplest possible example of a TR-symmetric type-II WSM. 

\begin{figure}
\includegraphics[height=1.85in]{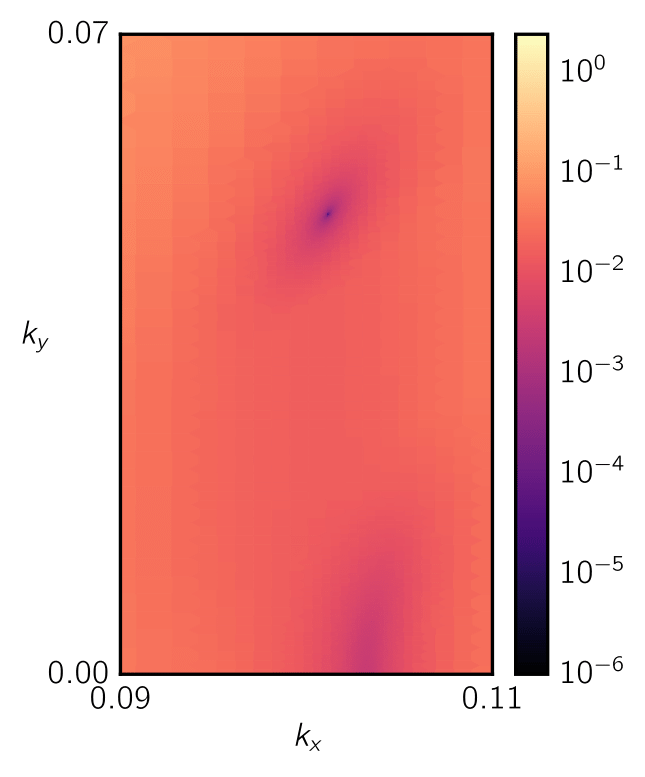}
\caption{The gap between bands $N$ and $N+1$ in MoTe$_2$ at $k_z=0$. The gap is small but non-zero on the $k_y=0$ line, signaling a region, where additional Weyl points can arise. The point of zero gap at $k_y=0.05$ is the Weyl point $W$ described in the main text.}
\label{fig:gap}
\end{figure}

\begin{figure}
\includegraphics[width=\columnwidth]{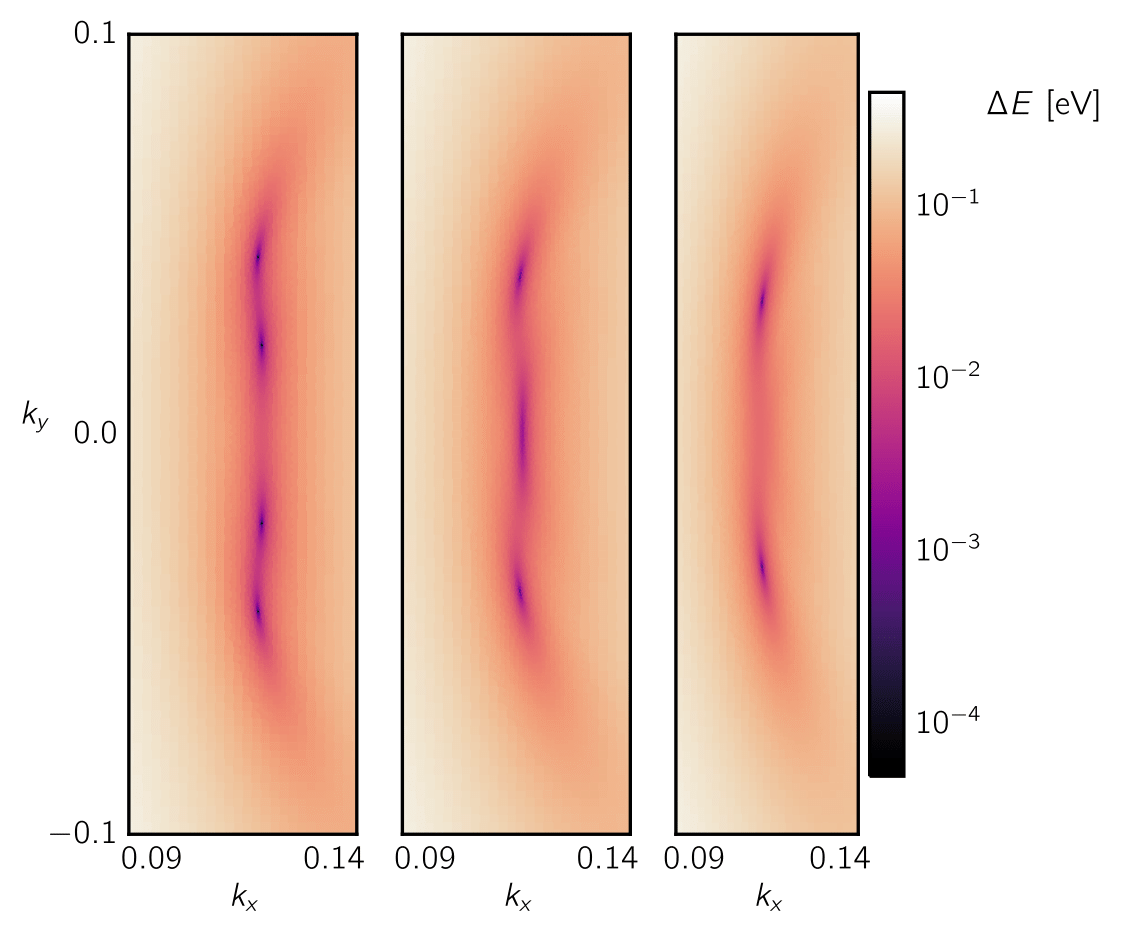}
\caption{Schematic illustration of the possible topological phase transition between the states of 4 and 8 type-II Weyl points which can be driven by strain.}
\label{newWP}
\end{figure}

The previous study of Ref.~\cite{Sun-PRB15} only considered crossings of bands $N$ and $N+1$ and reported 8 type-II Weyl points in MoTe$_2$ in the $k_z=0$ plane. This discrepancy arises due to the difference in the lattice parameters. Although the difference is small, the 100~K crystal structure reported here is close to the topological phase transition point, where additional Weyl points appear in pairs of opposite chirality from the $k_y=0$ mirror plane. This can be seen in the Fig.~\ref{fig:gap}, where the gap between the bands $N$ and $N+1$ is plotted in the region of interest for $k_z=0$ cut of the Brillouin zone.  
A minor change in the lattice constants can give rise to 4 additional type-II Weyl points in the $k_z=0$ plane, as illustrated schematically in Fig.~\ref{newWP}. 
Extreme sensitivity of band structure topology in MoTe$_2$ and the fact that the results of Ref.~\cite{Sun-PRB15} are obtained for the  different temperature structure suggest the possibility of temperature-driven topological phase transitions in this material.

\section{$\mathbb Z_2$ Invariants and their Relation to Weyl points}
\label{app:z2index}

\begin{figure}[tb]
\includegraphics[width=0.48\columnwidth]{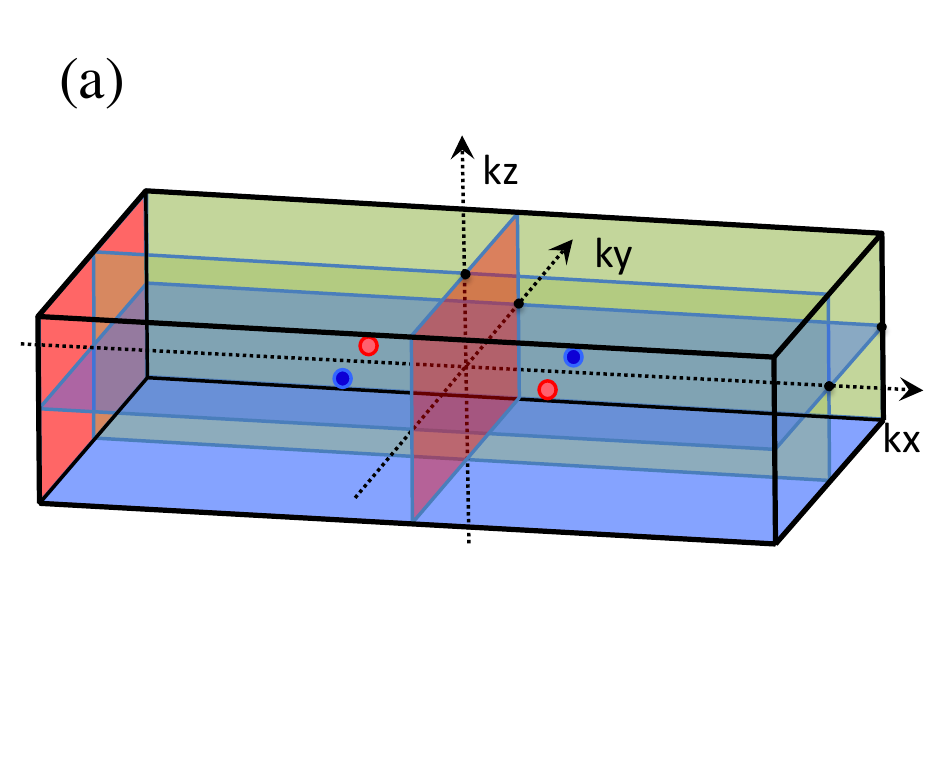}
\includegraphics[width=0.48\columnwidth]{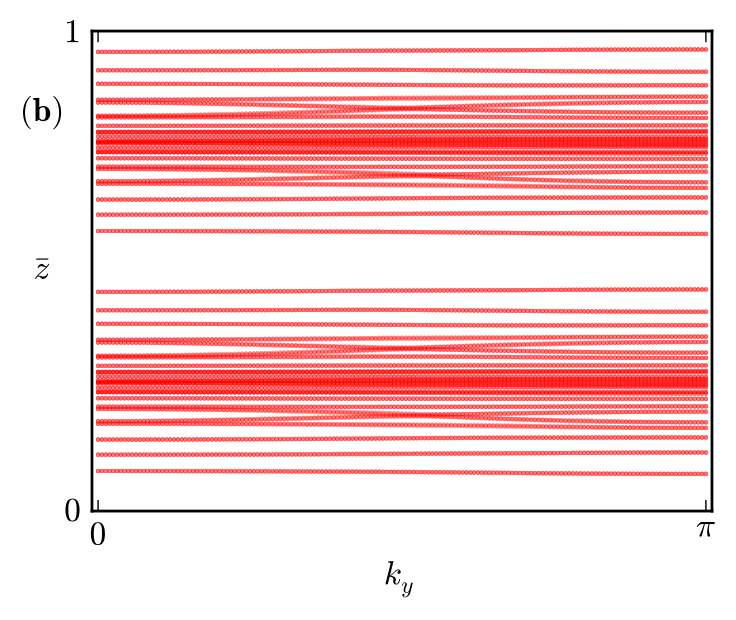}
\includegraphics[width=0.48\columnwidth]{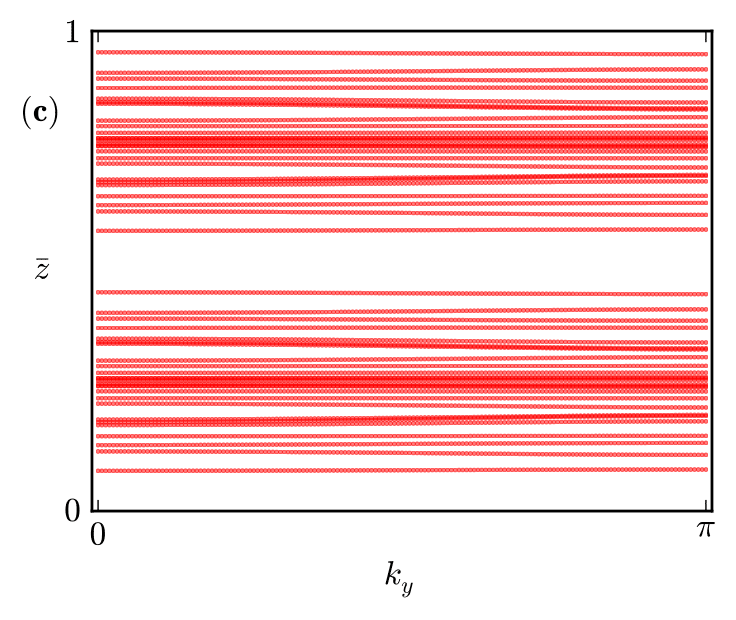}
\includegraphics[width=0.48\columnwidth]{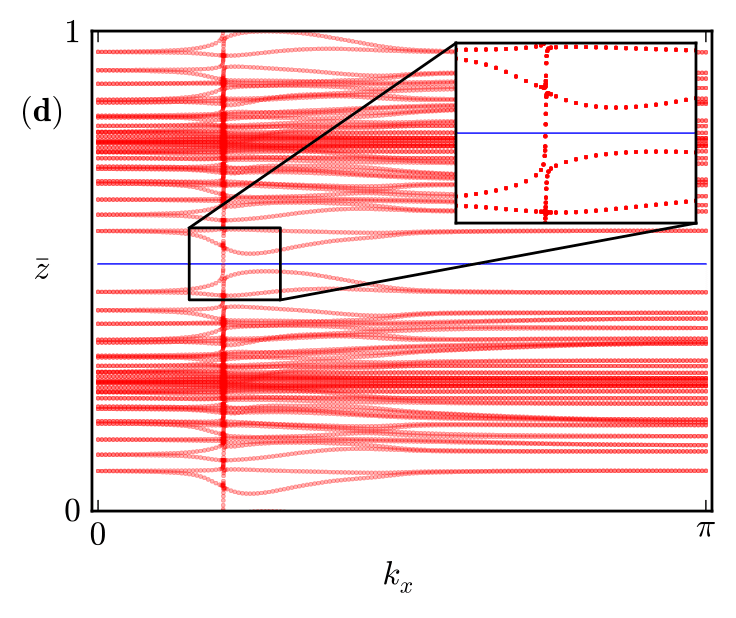}
\includegraphics[width=0.48\columnwidth]{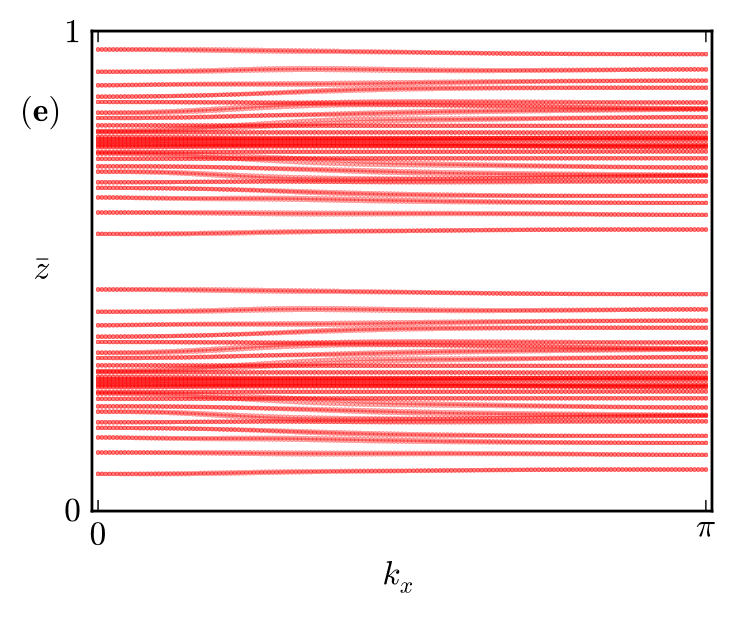}
\includegraphics[width=0.48\columnwidth]{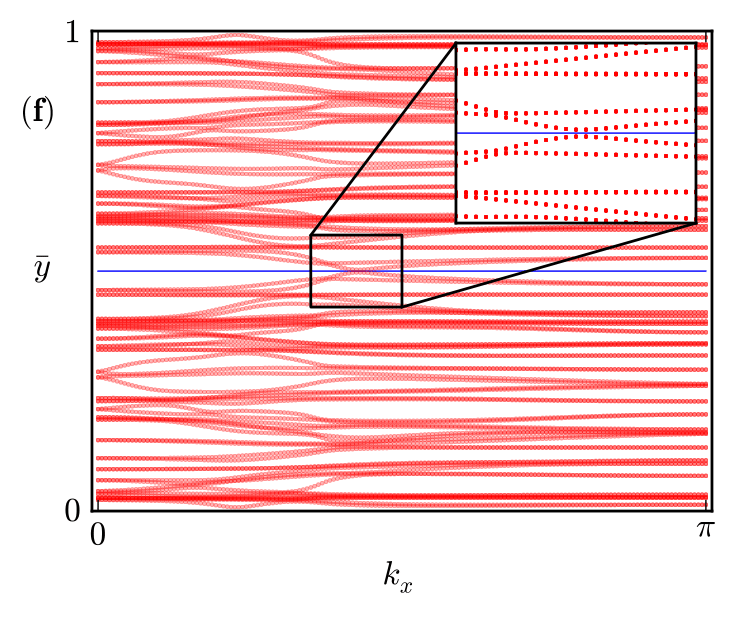}
\caption{Upper left panel: Six time-reversal symmetric planes are shown in the Brillouin zone. The $k_i=0,0.5$ planes are denoted by the red-, green-, and blue-colored sheets, for $i=x,y$ and $z$, respectively. Other panels show the flow of Wannier charge centers for the five gapped planes. Panel (b): $k_x = 0$; panel (c): $k_x = \pi$; panel (d): $k_y = 0$; panel (e): $k_y = \pi$; panel (f): $k_z = \pi$. The $\mathbb Z_2$ invariant is well defined on all except the $k_z=0$ plane, and is non-trivial only on $k_y=0$ plane.}
\label{six}
\end{figure}

We establish a connection between $\mathbb{Z}_2$ invariants used for insulators and the existence of Weyl points.  If the usual $\mathbb{Z}_2$ invariant is nonzero only on one out of the $6$ common high-symmetry time-reversal symmetric (TR) planes ($k_i=0$ and $k_i=0.5$), the system has to exhibit Weyl points. 
This is easy to see. Consider the TR planes shown in the upper left panel of Fig.~\ref{six}. Let only one of the planes be $\mathbb{Z}_2$ non-trivial, so that it exhibits a quantum spin Hall effect. The edge modes of the quantum spin Hall effect on this plane can result from (a) a closed surface Fermi surface such as in a weak or strong topological insulator or (b) from a disconnected open surface Fermi surface such as the Fermi arcs. Case (a), however, would imply the existence of another nontrivial $\mathbb{Z}_2$ index on one of the other TR planes (either parallel or perpendicular to the non-trivial one), hence the only possibility is that of an open Fermi arc surface state.  This exercise also reveals the canonical connection pattern between Weyl points on the surface: the Fermi arcs will form the continuation of the quantum spin Hall edge states off the high-symmetry plane. 


In MoTe$_2$ the valence bands and the conduction bands ($N$ and $N+1$th bands) are directly gapped on five out of six TR planes, with the exception of the $k_z=0$ plane that hosts four Weyl points. The appearance of Weyl points and the connection of Fermi arcs can  be deduced by analyzing the $\mathbb Z_2$ invariants~\cite{Kane-PRL05-b} for the these five TR planes. Fig.~\ref{six} shows the flow of Wannier centers~\cite{Soluyanov-PRB11-a, Soluyanov-PRB11-b, Yu-PRB11} on the five planes as calculated directly from first-principles calculations~\cite{Z2Pack}. All but the $k_y=0$ planes are $\mathbb{Z}_2$ trivial, so that the quantum spin Hall effect appears only in the $xz$-plane, guaranteeing that non-trivial surface states exist and cross the $k_x$ axis on the $(001)$ surface in accord with the surface state calculation illustrated in Fig.~\ref{arc} below.

\section{Topological charge of Weyl points}

Using the crystal symmetry $C_{2v}$, we only need to calculate the topological charge of the Weyl points within one-fourth of the entire Brillouin zone.  The topological charge $(C_S)$ of a Weyl point  can be defined as the net flux of the Berry gauge field penetrating a 2D surface~\cite{Wan-PRB11,wang1, Soluyanov-ARX15}
\beq
C_S=\frac{1}{2\pi}\oint_S[\nabla_{\bf k} \times{\bf  A(k)}]\cdot d{\bf S}
\label{eqcs}
\eneq
where the integrand ${\bf A(k)}=-i\langle u_{\bf k}|\nabla_{\bf k}|u_{\bf k}\rangle$ is the Berry connection for the Bloch states $|u_{\bf k}\rangle$ calculated on the surface $S$ that encloses the Weyl node. By Stokes theorem, the $C_S$ defined above should be equal the topological charge of the Weyl point. 

For this reason, the closed Fermi surface of the type-I Weyl point has nonzero topological charge. In the case of a type-II Weyl point, however, the Fermi surface is open, and hence cannot be used to compute the topological charge of the Weyl point. Instead, we integrate the Berry curvature computed for $N$ lowest bands, where $N$ is the number of electrons per unit cell. A closed surface, on which the lowest $N$ states are separated by an energy gap from the other higher energy states, and which encloses the type-II Weyl, can easily be found. This surface defines a 2D manifold in 3D momentum space, formed by the lowest $N$ states, and unlike any possible Fermi surface, it corresponds to different energy values for different momenta ${\bf k}$ in the Brillouin zone. 

\begin{figure}[tb]
\centering
\includegraphics[width=8 cm]{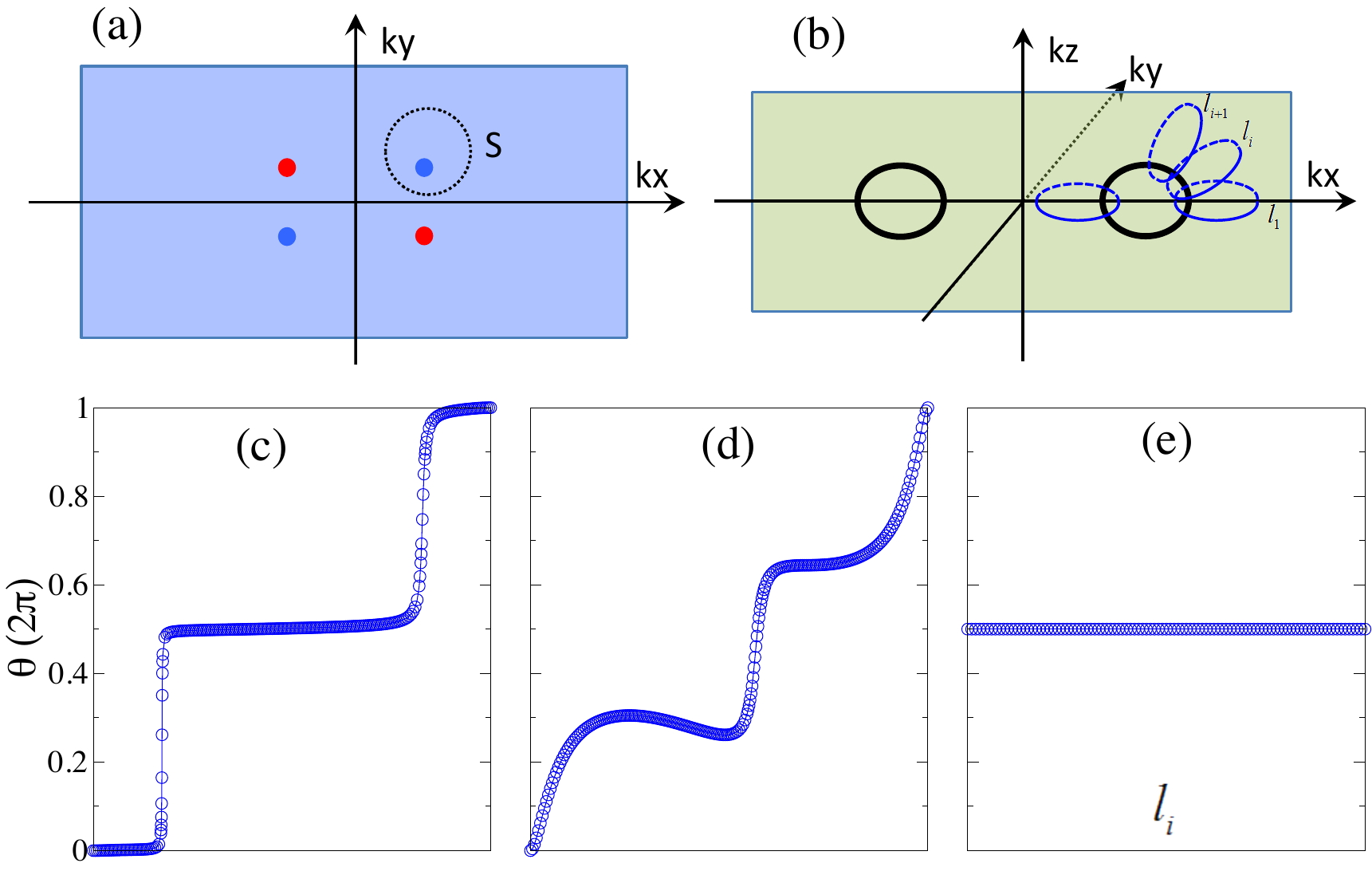}
\caption{(a) A sphere surrounding one of the $W$ points is shown schematically. (b) Loops linking with the nodal line formed by bands $N-1$ and $N$ are shown. (c) Topological charge for the point $W$. (d) topological charge of one of the Weyl points formed by bands $N+1$ and $N+2$, obtained by integrating the Berry curvature of $N+1$ bands. (e) The Berry phase acquired by Bloch states when going around a gapped loop linking with the nodal ring. For all loops this phase is equal to $\pi$, as expected.}
\label{wlp}
\end{figure}

Using both, first-principles calculations and Wannier-based tight-binding models~\cite{Souza-PRB01, wannier90}, Bloch states were calculated on spheres enclosing Weyl points. One of such spheres enclosing the point $W$ of the main text is shown as a circle $S$ in the $k_z=0$ plane of the Brillouin zone in Fig.~\ref{wlp}(a). For the calculation of the topological charge of Weyl points formed by bands $N+1$ and $N+2$ the Berry curvature is computed for $N+1$ bands and the integration surface is chosen such that an energy gap between $N+1$ and $N+2$ bands is present everywhere on it. For further illustration of the topological charge, following the work of Ref.~\cite{Soluyanov-ARX15} in Fig.~\ref{wlp}(c-d) we plot the total electronic polarization~\cite{King-Smith-PRB93} for one-dimensional circular cuts of the sphere $S$ taken for different values of the polar angle $\theta$ for the Weyl points $W$ and one of the points formed between bands $N+1$ and $N+2$ located at $(0.1004,0.040,0.0)$.  The shift of polarization value when going from $\theta=0$ to $\theta=\pi$ gives the Chern number (chirality) of the Weyl point.

To prove the existence of the line nodes formed by bands $N-1$ and $N$, we calculated the Berry phase acquired by $N-1$ bands along a loop in $k$-space linked with one of the nodal lines, as shown if Fig.~\ref{wlp}(b). On each of such loops a gap between bands $N$ and $N-1$ exists, so that the Berry phase for the manifold of $N-1$ bands is a well-defined quantity. In the presence of a monopole inside the loop, this Berry phase has to be equal to $\pi$. As illustrated in Fig.~\ref{wlp}(c), all loops have $\pi$ Berry phase thus proving the existence of the line node (that is, a monopole exists within every loop).

Pierced by a line on which the bands $N-1$ and $N$ are degenerate, while being gapped on the loop and due to the presence of the mirror plane this Berry phase is equal to $\pi$ as illustrated for a set of loops in Fig.~\ref{wlp}(e).  

\subsection{Weyls and Nodal Lines formed by bands other than $N$ and $N+1$}

We found a plethora of topological features formed by bands $N+1$ and $N+2$, including line nodes on the mirror planes and several sets of Weyl points. Of these the ones found at ${\bf k_1}=(0.1004,0.040,0.0)$, ${\bf k_2}=(0.11307861, 0.06131836, 0)$, ${\bf k_3}=(0.1603, 0.0750, 0)$, and ${\bf k_4}=(0.1196, 0.1068, 0.2508)$ (and their mirror images) are of the most relevant located only $60$, $57$, $73$ and $66$~meV above the Fermi level. Hidden inside the carrier pockets these additional Weyl points and their associated Fermi arcs overlap with the bulk states when projectod onto the experimentally relevant $(001)$-surface. The nodal lines present on the $k_x=0$ and $k_y=0$ planes (including the one formed by bands $N-1$ and $N$) also do not contribute visible spectroscopic signatures to this surface -- their associated drum-head surface states~\cite{Yu-PRL15} are projected onto the surfaces other than $(001)$.

\begin{figure*}[t]
\centering
\includegraphics[width=16 cm]{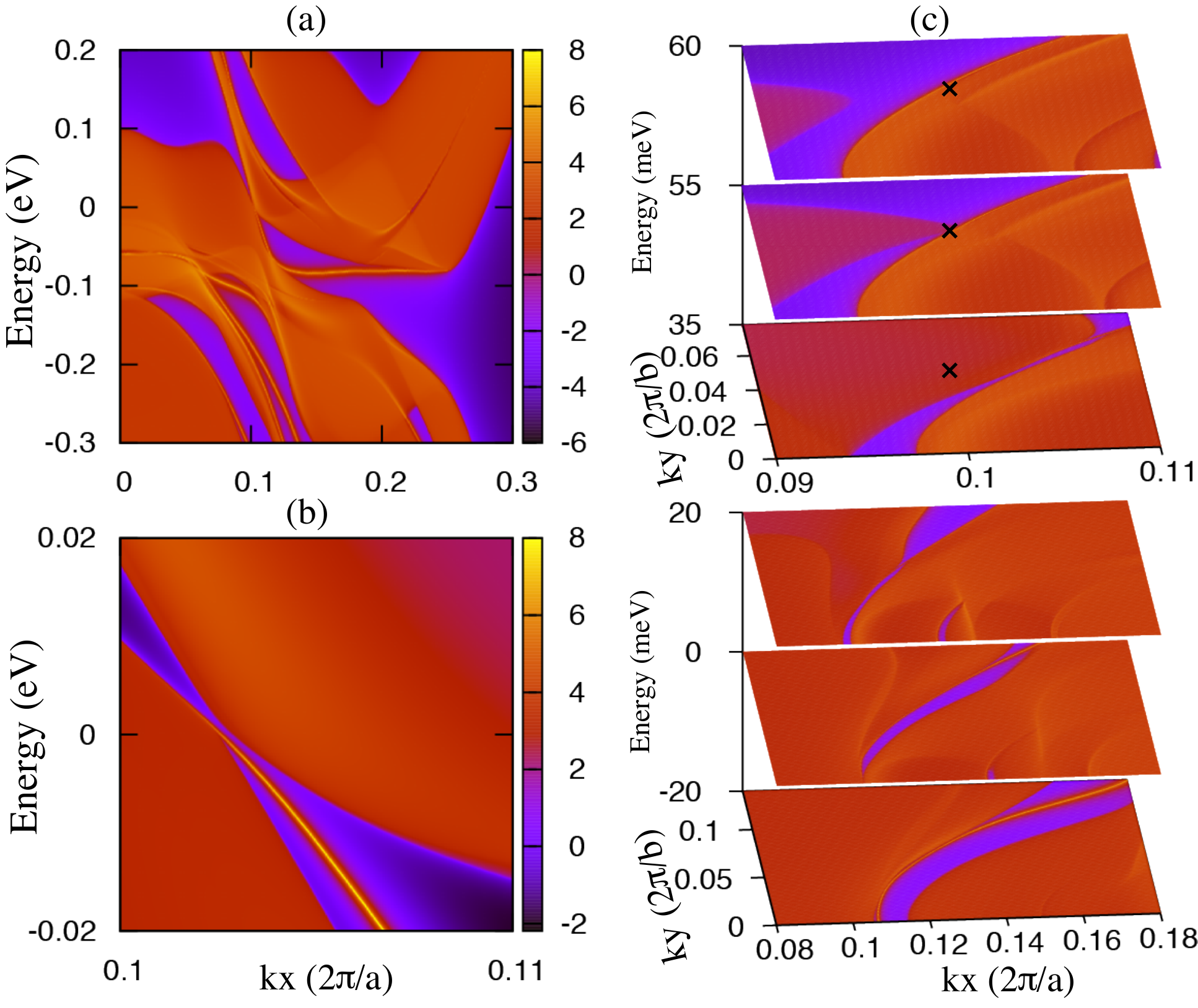}
\caption{
Surface dispersions and Fermi surfaces.
(a): $(001)$-surface dispersion in the $k_x$ direction. (b): Zoom-in of panel (a), showing in-gap topological surface states. (c): The surface Fermi surfaces and Fermi arcs for different values of the chemical potential. 
} \label{arc}
\end{figure*}

\begin{figure*}[t]
\centering
\includegraphics[width=16 cm]{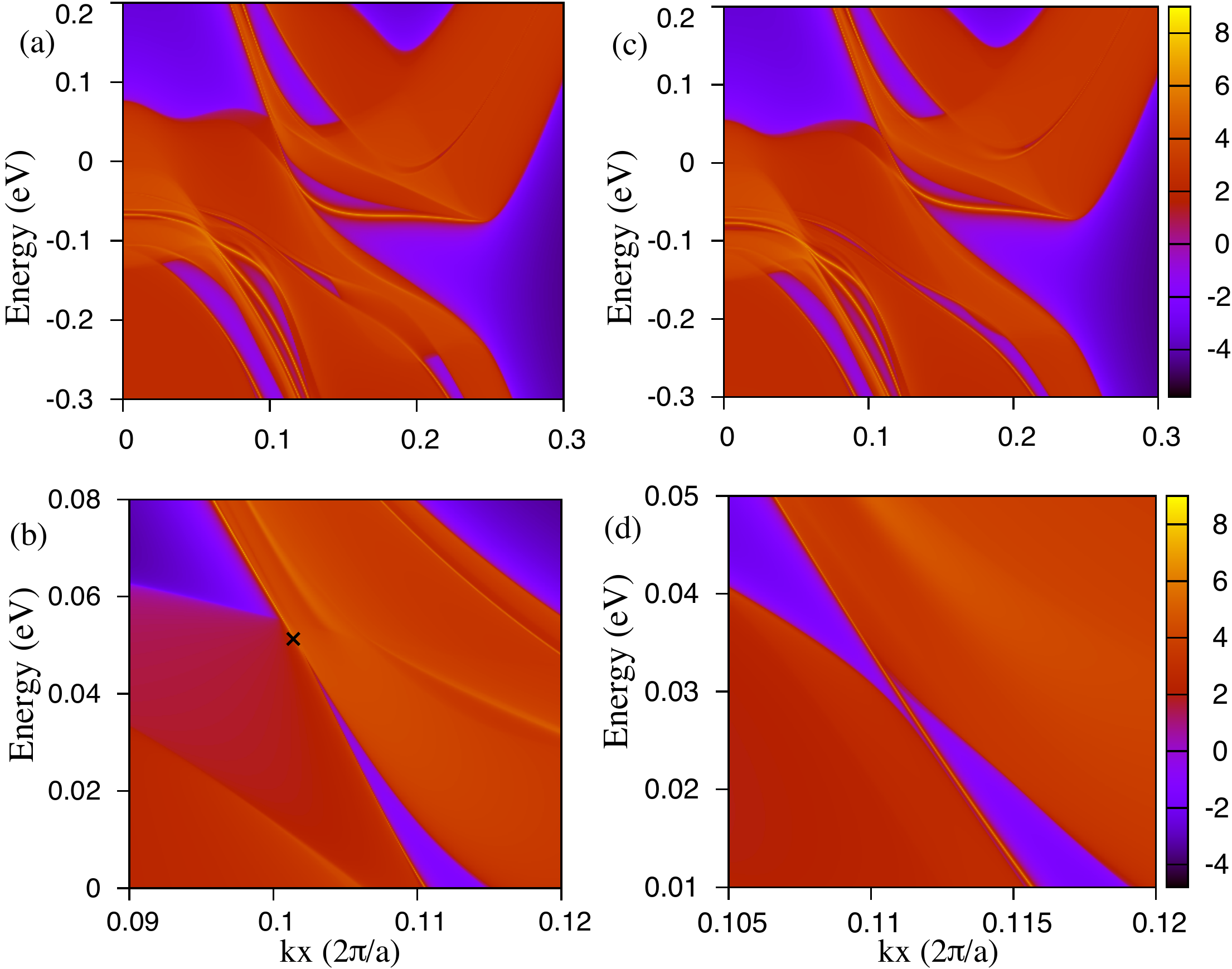}
\caption{Surface dispersions in the $k_x$ direction with different $k_y$-values.
(a): $(001)$-surface dispersion in the $k_x$ direction with $k_y=0.0503$. (b): Zoom-in of panel (a), showing a gap closure at the $W$ Weyl point. 
(c): $(001)$-surface dispersion in the $k_x$ direction with $k_y=0.07$. (d): Zoom-in of panel (c), showing a trivial surface state. 
} \label{arc2}
\end{figure*}

\section{Fermi arcs and surface states}

Implementing the Green's function method of Ref.~\cite{Sancho-JPhysF85} to the Wannier-based tight-binding Hamiltonian generalted from the first-principles calculation, we obtained the surface states for the $(001)$-surface. They are plotted in Fig.~\ref{arc}, together with the corresponding surface Fermi surfaces. As discussed above, the Weyl points formed by bands $N$ and $N+1$ are responsible for the presence of visible surface states in the bulk gapped region. In accord with the discussion above, the surface states connecting valence and conduction states along the $k_x$ direction are clearly seen in the spectral function of the $(001)$-surface of Fig.~\ref{arc}(a-b). 

The connectivity patterns of the corresponding Fermi arcs for the $(001)$-surface at different energies are shown in Fig.~\ref{arc}(c). Since MoTe$_2$ is a type-II Weyl semimetal, the Fermi arcs are always accompanied by the projections of bulk electron and hole pockets. This makes it possible to tune the Fermi arcs by changing the position of the chemical potential, and Fig.~\ref{arc}(c) illustrates the evolution of the Fermi arc states for different values of the chemical potential. The states are clearly visible when this value is set to $-20$~meV below the Fermi level. 

Unlike the case of type-I Weyl semimetals, where the Fermi arc necessarily connects the projections of the Weyl points onto the Fermi surface, here it arises out of the generic point in the electron pocket and dives back into it, as illustrated in the main text. This arc, however is still topologically non-trivial, as illustrated in Fig.~\ref{arc2}. At small values of $|k_y|<0.0503$ in between the two $W$ points the 2D $(k_x,k_z)$ cut of the Brillouin zone exhibits the quantum spin Hall effect, and the corresponding topological surface state is clearly visible connecting the valence and conduction bands across the gap (see Fig.~\ref{arc}(a-b)). At the position of $W$ points, that is for $k_y=\pm 0.0503$ the corresponding 2D cut of the Brillouin zone is metallic but a surface state is still seen below the Weyl point at the boundary of the bulk valence bands projection (Fig.~\ref{arc2}(a-b)), but it now reconnects from the valence to conduction states. At $|k_y|>0.0503$ the 2D cuts of the Brillouin zone become topologically trivial and a topologically trivial surface state is clearly seen in Fig.~\ref{arc2}(c-d). Thus, the Fermi arc of the main text is formed by a topological surface state, resulting from the quantum spin Hall effect at small $k_y$, and the topologically trivial state at larger $|k_y|$ serves to connect this in-gap state to the projections of the bulk Fermi pockets.

\section{Strains}

As mentioned in the main text, the band structure of MoTe$_2$ around the Fermi level is very sensitive to changes in the lattice constant. To illustrate this we studied topological phase transitions occurring between the valence and conduction bands in this material under various strain values (see Supplementary Information).  We find that for these bands two additional sets of WPs can appear in MoTe$_2$ with small changes in the lattice constants.  The first set consists of 4 type-II WPs in the $k_z=0$ plane arising in pairs of opposite chirality from the mirror plane $k_y=0$ and giving 8 total WPs in analogy with WTe$_2$~\cite{Soluyanov-ARX15}. This scenario is realized under a hydrostatic strain of~\footnote{Approximate values of strains are given.} $+0.3$\% and uniaxial strains of $+2$\% and  $-0.3$\% in $z$ and $y$ correspondingly. Another set of additional WPs consists of type-II nodes appearing off the $k_z=0$ plane for a hydrostatic strain of $-0.25$\% and for uniaxial strains of $-0.2$\% and $+2$\% in $z$ and $y$ correspondingly. Finally, both sets appear for a uniaxial strain in the $x$ direction of $0.5$\% (only $0.1$\% strain is required to generate the additional set at $k_z=0$), while for negative strains in $x$ no new WPs are generated between bands $N$ and $N+1$, but the $W$ points move closer to each other~\footnote{For a strain of $-1$\% the $W$ point moves to $(0.09695, 0.03356, 0)$.}. The strong dependence of the Weyl physics on the applied strain has also been pointed out in ~\cite{Sun-PRB15}. For their structure, it was also found that strain can induce a type-II to type-I Weyl transition ~\cite{Sun-PRB15}.  
The calculation of strained structures has been performed as follows. The aim of this study was to look at the stability of different topological phases under small changes in the lattice parameters. Here the strained structures were calculated from first-principles by changing the lattice constants only, without further relaxation. To find the critical values of strains at which the topological phase transitions occur, {\it ab initio} calculations were performed for a large number of strained structures and the closing of the band gap at the Weyl nodes was observed. The full analysis including the calculation of topological invariants, however, was done only for specific strain values to confirm the qualitative nature of the phase transitions. The approximate values of critical strains were extracted from these calculations.

The details on the calculation of strained structures are as follows.  The aim of this study was to look at the stability of different topological phases under small changes in the lattice parameters. Here the strained structures were calculated from first-principles by changing the lattice constants only, without further relaxation. To find the critical values of strains at which the topological phase transitions occur, {\it ab initio} calculations were performed for a large number of strained structures and the closing of the band gap at the Weyl nodes was observed. The full analysis including the calculation of topological invariants, however, was done only for specific strain values to confirm the qualitative nature of the phase transitions. The approximate values of critical strains were extracted from these calculations.
\section{Additional remarks}
VESTA~\cite{vesta}, Gnuplot~\cite{gnuplot}, Mayavi~\cite{mayavi}, Matplotlib~\cite{matplotlib}, GIMP~\cite{gimp} and Inkscape~\cite{inkscape} software packages were used to prepare some of the illustrations.

\end{document}